\documentclass[iop]{emulateapj}
\usepackage{multirow}
\usepackage{natbib}
\usepackage{epstopdf}
\begin{document}

\title{Using Red Clump Stars to Decompose the Galactic Magnetic Field with Distance}
\author{Michael D. Pavel}
\affil{Department of Astronomy, The University of Texas at Austin, 2515 Speedway, C1402, Austin, TX 78712 USA}
\email{pavelmi@astro.as.utexas.edu}

\slugcomment{Accepted for Publication in the Astronomical Journal}

\shorttitle{Decomposing Magnetic Fields}
\shortauthors{Pavel, M. D.}

\begin{abstract}

A new method for measuring the large-scale structure of the Galactic magnetic field is presented. The Galactic magnetic field has been probed through the Galactic disk with near-infrared starlight polarimetry, however the distance to each background star is unknown. Using red clump stars as near-infrared standard candles, this work presents the first attempt to decompose the line of sight structure of the sky-projected Galactic magnetic field. Two example lines-of-sight are decomposed: toward a field with many red clump stars and toward a field with few red clump stars. A continuous estimate of magnetic field orientation over several kiloparsecs of distance is possible in the field with many red clump stars, while only discrete estimates are possible in the sparse example. toward the Outer Galaxy, there is a continuous field orientation with distance that shows evidence of perturbation by the Galactic warp. toward the Inner Galaxy, evidence for a large-scale change in the magnetic field geometry is consistent with models of magnetic field reversals, independently derived from Faraday rotation studies. A photo-polarimetric method for identifying candidate intrinsically polarized stars is also presented. The future application of this method to large regions of the sky will begin the process of mapping the Galactic magnetic field in a way never before possible.

\end{abstract}

\keywords{Infrared: ISM --- ISM: magnetic fields --- stars: distances --- stars: horizontal-branch --- Techniques: photometric --- Techniques: polarimetric}

\section{Introduction}

Observationally, our knowledge of the large-scale structure of the Galactic magnetic field has primarily been derived from studies of background starlight polarimetry \citep{1970MmRAS..74..139M, 1996ApJ...462..316H, 2000AJ....119..923H, 2012ApJ...749...71P}, Faraday rotation \citep{1997AA...322...98H, 2006ApJ...642..868H, 2007ApJ...663..258B, 2010ApJ...714.1170M, 2012ApJ...755...21M, 2011ApJ...738..192P, 2011ApJ...728...97V}, and synchrotron emission \citep{2012ApJ...757...14J, 2012ApJ...761L..11J}. Yet only Faraday rotation has successfully resolved large-scale magnetic field structures with distance. This work presents a new method for resolving the large-scale magnetic field geometry with distance by using red clump stars as standard candles in the Galaxy. Combined with near-infrared (NIR) starlight polarimetry, the magnetic field properties of a line of sight can be decomposed with distance to map variations in the plane-of-the-sky magnetic field.

Previous Faraday rotation studies have applied similar methods to studies of the large-scale structure of the Galactic magnetic field. By combining observations of Galactic pulsars with distance estimates (that sample discrete portions of the Galactic magnetic field) with observations of polarized extragalactic sources (that sample the entire Galactic line of sight towardh that object) detailed models of the Galactic magnetic field have been generated \citep[e.g., ][]{2007ApJ...663..258B, 2008AA...477..573S, 2011ApJ...728...97V}. However, these studies suffer from the limitations of Faraday rotation, specifically that they only sense the line of sight component of magnetic fields. Also, pulsar distances can be uncertain. Few parallax measurements are available and only for nearby pulsars, H{\scriptsize I} kinematic distances typically only give upper or lower distance limits, and those distances have typical uncertainties of 0.5 kpc or greater \citep{1990AJ....100..743F, 2012ApJ...755...39V}.

Starlight polarization arises as an integral quantity of the line of sight between the background star and the observer. Regions of magnetically-aligned dust grains impose weak polarizations parallel to the local magnetic field direction  on background starlight. Via this mechanism the orientation of magnetic fields are directly measurable \citep{1985ApJ...290..211L, 2011ApJ...740...21P}, the plane-of-sky magnetic field strength can be indirectly inferred \citep{1953ApJ...118..113C, 2008ApJ...679..537F, 2012ApJ...755..130M}, though the field direction is not obtainable. Statistical studies of observed starlight polarization have shed light on the general symmetry and spiral-type pitch angle of the Galactic magnetic field \citep{1996ApJ...462..316H, 2012ApJ...749...71P}, but even kpc-scale details of the morphology of the magnetic field remain hidden. To measure changes in the sky-projected magnetic field with distance, as is done with Faraday rotation for the line of sight field, reliable stellar distance markers are needed.

The first steps in using starlight polarimetry to disentangle magnetic fields with distance have already been taken. \citet{2009ApJ...690.1648N} attempted to measure magnetic fields in the Galactic Center with NIR polarimetry by subtracting the polarization associated with foreground ``bluer'' field stars from the polarization of ``redder'' stars located near the Galactic Center and showed that both populations were polarimetrically indistinguishable. This work was followed by \citet{2010ApJ...722L..23N} who used the same method to observe a transition from toroidal to poloidal magnetic field above and below the Galactic center. \citet{2013AJ....145...74C} use a similar foreground subtraction method to remove a foreground Galactic contribution and resolve the magnetic field of a galaxy seen through the Taurus molecular cloud. This work extends magnetic field decomposition by moving away from statistical treatments of foreground stars, and instead uses red clump stars as standard candles to identify and characterize changes in magnetic field orientation with distance.

In color-magnitude diagrams (CMDs) of star clusters, the red clump is an overdensity of core helium burning (horizontal branch) stars seen in intermediate-age ($\sim 1-10$ Gyr) clusters and old ($\geq 10$ Gyr), metal-rich clusters \citep{2000ApJ...539..732A}. The relative invariance of red clump star K-band ($2.2\mu$m) absolute magnitude with age or metallicity \citep[$M_K=-1.57\pm0.05$;][]{2007AA...463..559V} allows it to serve as a K-band standard candle \citep{1970MNRAS.150..111C, 2000ApJ...539..732A}. In a NIR CMD of field stars, red clump stars located at different distances form an easily distinguishable feature. The photometric properties and methods for identifying this feature have been studied in the Galactic disk \citep{2002AA...394..883L, 2012MNRAS.419.1637L, 2013ApSS.346...89K}, the Large Magellanic Cloud \citep{2007AJ....134..680G}, and in star clusters \citep{2009MNRAS.394L..74G, 2011AA...535A..33M}. These studies have further refined our knowledge of the intrinsic properties of these objects, and limitations to their use. 

This work implements the NIR red clump selection algorithm described in \citet{2002AA...394..883L} to identify red clump stars at different distances. For a population of stars with a fixed absolute magnitude, the apparent magnitude is an excellent proxy for distance order, since extinction monotonically increases with distance along a given line of sight. Naively assuming a constant dust density along a sightline, the physical distances to these stars are estimated below. These stars are then used as distance markers along a line of sight to probe magnetic field properties between them via NIR starlight polarimetry.

The details of a new method for measuring the Galactic magnetic field structure with distance is presented in Section 2. The method is applied to example lines-of-sight toward the inner Galaxy in Section 3 and the outer Galaxy in Section 4. Section 5 presents new results on the morphology of the Galactic magnetic field toward these two lines-of-sight and also discusses evidence for intrinsically polarized red clump stars. A summary and conclusions are presented in Section 7.

\section{Decomposing Sky-Projected Magnetic Field Orientation Along the line of sight}

Starlight linear polarization arises from dichroic extinction via aligned dust grains along the line of sight between a star and an observer. Spinning, non-spherical dust grains preferentially align with their long axes perpendicular to the local magnetic field orientation \citep{2003JQSRT..79..881L, 2007JQSRT.106..225L, 2011ASPC..449..116L}. These aligned dust grains preferentially extinct photons polarized perpendicular to the local magnetic field direction (parallel to a grain's long axis), giving rise to a weak linear polarization along the magnetic field direction. \citet{2011ApJ...740...21P} describes the equations of polarized radiative transfer and discusses the effects of magnetic field geometry on the observed polarization properties.

\begin{figure}
	\epsscale{1.15}
	\centering
		\plotone{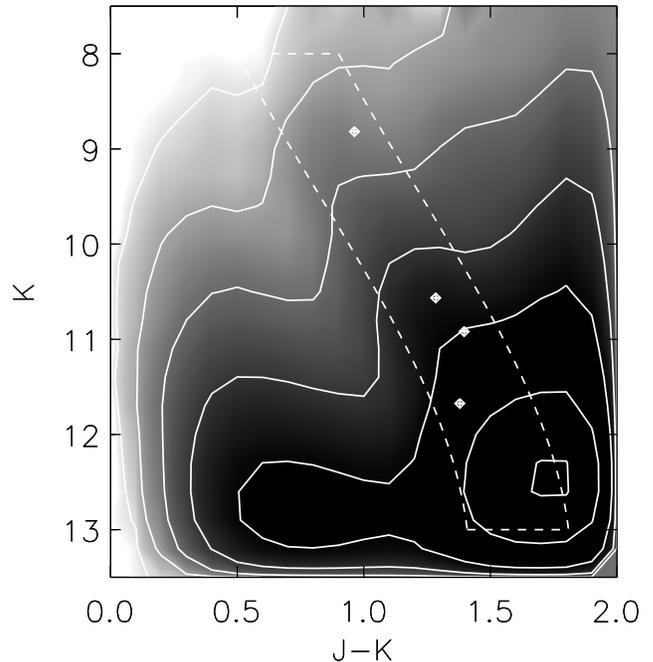}
	\caption{\label{cmd}Near-infrared color-magnitude diagram showing field stars toward a $100' \times 100'$ region centered on $(\ell , \it{b})$ = (19.49\degr, +0.56\degr), from 2MASS. White contours and the background greyscale image show the relative density of stars. The red clump sequence found using the method of \citet{2002AA...394..883L} is shown by the dashed white lines. The solid white triangle approximately traces the main sequence. The locations of red clump stars with $H$-band polarimetry in the $10' \times 10'$ centered on the same coordinates are identified in the figure.}
\end{figure}

Summarizing \citet{2011ApJ...740...21P}, linear polarization will trace the average, extinction-weighted, sky projected orientation of the magnetic field between an unpolarized star and observer. Magnetic field components parallel to the line of sight do not contribute to this polarization, though these are the fields traced by Faraday rotation. The Galactic magnetic field consists of regular and turbulent components in approximate equipartition. Along long (kpc) lines-of-sight in the diffuse interstellar medium, the turbulent magnetic field contributions will tend to cancel, allowing linear polarimetry to trace the large-scale structure of the Galactic magnetic field.

If the distances to stars with measured polarizations can be estimated, then the average sky-projected magnetic field orientation in the material between those stars can be measured via differences in their Stokes parameters. These differences arise in the dusty, magnetized medium between the stars. This method of decomposition requires that the different stars sample the same line of sight so that polarizing foregrounds can be reliably removed. If the angular separation on the sky between stars is large, then their light may not be sampling the same polarizing layers. The typical size scales for significant variation in the random component of the Galactic magnetic field are $10-100$ pc \citep{1993MNRAS.262..953O, 2008ApJ...680..362H}. Two stars 5 kpc distant and separated by 10 arcmin on the sky would only be physically separated by 14.5 pc, which is smaller than the typical random fluctuations in the Galactic magnetic field, and much smaller than the size scales for variations in the regular magnetic field. As the light from these two stars approach an observer and converge, they propagate through almost identical paths. If instead one star is located between the observer and the other star, the conditions are favorable for measurement of the magnetic properties between the stars via Stokes subtraction.

\subsection{Identifying Red Clump Stars}
\label{red_clump}

Red clump stars are identified using the CMD method described in \citet{2002AA...394..883L}, illustrated in Figure \ref{cmd} with Two Micron All Sky Survey (2MASS) data \citep{2006AJ....131.1163S} for a $100' \times 100'$ region centered on $(\ell , b)= (19.49\degr,+0.56\degr)$. In Figure \ref{cmd}, two significant features are seen. Main sequence stars of various reddenings are located in the region $0.5 < (J-K) < 1.0$ and $K > 8 + 4(J-K)$ (approximately traced by the solid white triangle in Figure \ref{cmd}), and the roughly diagonal feature running from the top-center toward the lower-right is dominated by red clump stars.

To objectively quantify the stars in the red clump CMD sequence, \citet{2002AA...394..883L} first grouped all stars by $K$ in 0.5 mag wide groups for $8<K<13$. For each $K$ group, a histogram of $(J-K)$ was calculated with 0.05 mag wide $(J-K)$ bins. This histogram was fit with a Gaussian to identify the $(J-K)$ location of the peak of the red clump feature at that $K$. The locus of points tracing the red clump stars was fit with a polynomial. Stars within 0.2 mag of the $(J-K)$ center of the feature are considered red clump stars. In Figure \ref{cmd}, a dashed line outlines this region.

\begin{figure}
	\epsscale{1.1}
	\centering
		\plotone{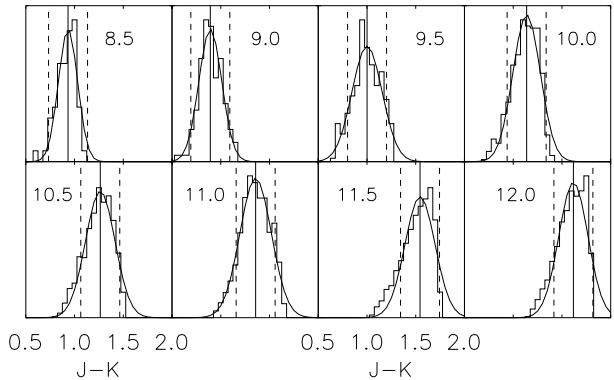}
	\caption{\label{gauss}Normalized histograms drawn from the CMD in Figure \ref{cmd} to illustrate the red clump fitting method. Each panel is labeled with the center of its $K$-band brightness. A Gaussian fit to each histogram is shown, the center of each fit is marked with a solid vertical line. The vertical dashed lines enclose $\pm0.2$ mag in $(J-K)$, the selection criteria for red clump stars.}
\end{figure}

The details of this fitting are shown in Figure \ref{gauss}. The well-separated red clump sequence in each $K$-band brightness bin (labeled in each panel) is shown. A similar histogram for the bluer main sequence is not shown in each brightness bin. The fitted Gaussian to each histogram is also shown. These centers of these fits are marked with vertical solid lines. Stars within 0.2 mag of each $(J-K)$ center are classified as red clump stars, and these color limits are shown by the vertical dashed lines in each panel. For fainter stars, the red end of this feature may suffer incompleteness (e.g., the lower right panel in Figure \ref{gauss}), which shifts the fitted center up to 0.15 mag bluer in $(J-K)$ than would be identified by eye. However, this remains a conservative method for identifying red clump stars. Some actual red clump stars are excluded so as to minimize false positive identifications. Sample contamination will be discussed in more detail in the next subsection.

The polarimetric observations described below cover single $10' \times 10'$ regions of sky, which typically contain too few stars with 2MASS photometry to reliably identify the red clump feature. To assist with identification of red clump stars within this small region, a much larger sample of 2MASS photometry is needed. As described above, a $100' \times 100'$ region was used for the red clump star identification. Then, only those red clump stars within the much smaller $10' \times 10'$ region were flagged for use. As an example, the four red clump stars identified in the $10'\times 10'$ region toward $(\ell , b)= (19.49\degr,+0.56\degr)$ are marked with diamonds in Figure \ref{cmd}

\subsection{Sample Contamination}

A key concern about the red clump identification method described above is the contamination of other stars into this region of the CMD. In particular, reddened main sequence stars could make up a significant fraction of the selected stars. \citet{2002AA...394..883L} comment on dwarf contamination and quantify the fraction of dwarfs in the total star counts as a function of magnitude (see Figure 5 in that paper). However, that analysis is for all stars in the field and does not quantify the effectiveness of the NIR color selection method at removing reddened interlopers.

\begin{figure}
	\epsscale{1.1}
	\centering
		\plotone{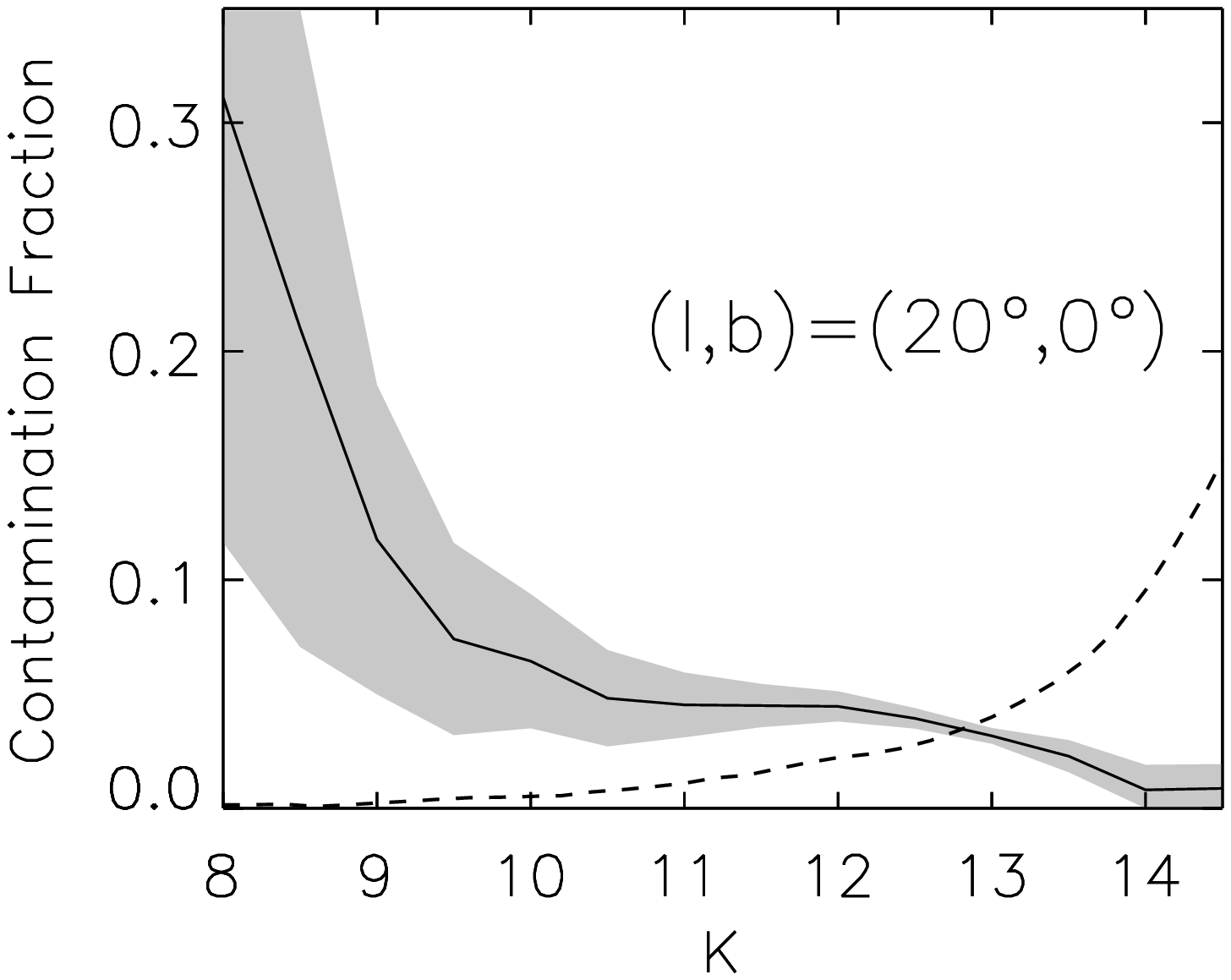}
	\caption{\label{contam}Two different measures of contamination toward the example direction $(\ell , \it{b})$ = (20\degr, 0\degr) using data from the stellar population synthesis model of \citet{2003A&A...409..523R}. The dashed line shows the fraction of dwarf stars at that magnitude. The solid line shows the dwarf and supergiant contamination fraction after applying the color selection criteria from \citet{2002AA...394..883L}. The shaded region estimates the combined intrinsic uncertainty in the contamination fraction and the uncertainty associated with varying the extinction per unit length toward this direction.}
\end{figure}

To estimate the total contamination along a line of sight, we use the stellar population synthesis model of \citet{2003A&A...409..523R} to generate sample star fields toward an example direction in the inner Galaxy, $(\ell , b)= (20\degr,0\degr)$. In addition to observed photometric properties, the intrinsic properties of each star are known. The average extinction per unit length was varied from $A_V /L = 0.7-1.8$ mag kpc$^{-1}$ in the different Monte Carlo samples. Using $A_V /L=$1.8 mag kpc$^{-1}$ is a reasonable assumption for the Galactic disk, but we also included lower values as a weak proxy for sightlines at higher Galactic latitudes.

The first quantity calculated was the dwarf fraction of all stars as a function of $K$-band apparent magnitude, the same quantity calculated in \citet{2002AA...394..883L} for the outer Galaxy. This quantity is plotted by the dashed line in Figure \ref{contam}. As shown in that figure, the dwarf contamination fraction rises for fainter stars. This contamination should be dominated by nearby, cool main sequence stars \citep[e.g., ][]{2007AJ....133..439C, 2008AJ....136.1778C, 2010AJ....139.2679B}. The contamination fractions calculated here are lower than the Outer Galaxy fractions from \citet{2002AA...394..883L}. This difference can be attributed to two effects: (1) lower extinction in the outer Galaxy allows intrinsically fainter stars to remain above a given brightness limit (reducing the Malmquist bias over a fixed volume), and (2) the distant (several kiloparsec) background in the outer Galaxy has a lower stellar space density of very bright sources (e.g., red clump stars) compared to the inner Galaxy.

Next, the contamination of the fitted red clump sequence was calculated. The photometric method described in Section \ref{red_clump} was applied to each of the simulated fields to identify red clump candidates. In this case dwarfs and supergiants, though rare, were considered as interlopers. The average fraction of interlopers is shown by the solid line in Figure \ref{contam} and the shaded region represents the uncertainty of that fraction. This uncertainty contains two components: the intrinsic uncertainty in the contamination fraction and the variation caused by different assumptions about the average extinction per unit length. In general, lower extinction per unit length is associated with more red clump contamination (toward the upper envelope of the shaded region), and larger extinction per unit length is associated with less contamination (toward the lower envelope of the shaded region). Later, the measured contamination fraction in Figure \ref{contam} will be used to estimate the probability of false detections.

\subsection{Distance Estimation}

To estimate the distance of a red clump star, the K-band absolute magnitude was assumed to be $M_K=-1.57\pm0.05$ \citep{2007AA...463..559V}. This is slightly dimmer than the value \citep[$M_K=-1.65$;][]{1992ApJS...83..111W,2000MNRAS.317L..45H} used by \citet{2002AA...394..883L}. K-band extinction was estimated from the apparent NIR (H-K) colors, assuming (H-K)$_o = 0.09\pm0.03$ \citep{2009BaltA..18...19S}, $R_V = 3.1$, and the \citet{1989ApJ...345..245C} extinction law.

\subsection{Isolating Polarizing Layers in the ISM}

Together, the position angle (PA) and degree of polarization ($P$) fully characterize the linear polarization properties of observed starlight, and correspond to a unique pair of linear Stokes parameters, $U$ and $Q$. Consider a series of $n$ optically thin polarizing layers characterized by Stokes parameters ($U_1$, $Q_1$), ($U_2$, $Q_2$),...,($U_n$, $Q_n$). The linear polarization ($U_f$, $Q_f$) of an initially unpolarized beam of light propagated through these layers becomes:
\begin{equation}
U_f = \displaystyle\sum\limits_{i=0}^n U_i
\end{equation}
\begin{equation}
Q_f = \displaystyle\sum\limits_{i=0}^n Q_i,
\end{equation}
under the assumption that all layers are optically thin. With measurements of ($U_f$, $Q_f$) through different numbers of polarizing layers, the properties of those polarizing layers can be measured by differencing Stokes parameters for stars at different distances. Similar techniques for the isolation of the polarization properties of stars in the Galactic center \citep{2009ApJ...690.1648N, 2010ApJ...722L..23N} and a background galaxy seen through Taurus \citep{2013AJ....145...74C} have been used before. However, those cases only considered the measurement and removal of a single, foreground polarization component.

In the following, this method is applied to the entire line of sight to decompose the sky-projected magnetic field orientation with distance. The first example below (Section \ref{inner}) demonstrates a coarse implementation of this procedure for an example line of sight with few identified red clump stars. In Section \ref{outer}, a more sophisticated implementation is shown later for a line of sight with many red clump identifications.

\subsection{The Effect of Discrete Clouds}

An individual (dense) molecular cloud along the line of sight probed by this method will have a significant effect on the observed starlight polarization. Since starlight polarimetry uses the {\em dust-weighted} magnetic field orientation, the presence of these higher density regions could contaminate the polarization signal expected from the large-scale Galactic magnetic field.

Comparative studies of sub-millimeter thermal dust polarization and optical starlight polarimetry have shown that the magnetic orientation of these clouds are generally aligned with the local magnetic field \citep{2009ApJ...704..891L, 2014arXiv1404.2024L}. This has also been observed in giant molecular clouds in M33 by \citet{2011Natur.479..499L}. The polarization signal superimposed by these clouds also depends on the degree of turbulence within the cloud: a magnetically turbulent cloud will produce a lower polarization signal per unit extinction than a less turbulent cloud \citep{2008ApJ...679..537F}.

However, these high-density regions can be identified via star counts \citep[e.g.,][]{2009ApJ...703...52L}, long wavelength continuum emission \citep[e.g.,][]{2011ApJS..192....4A}, or line emission from high-density tracers \citep[e.g., $^{12}$CO, $^{13}$CO, or CS;][]{2013ApJ...764..102F}. Velocity-resolved line emission at radio wavelengths is perhaps the best method for identifying individual high-density regions along a single line of sight. Once identified, the polarimetric effect of these clouds can be individually assessed. In the best case scenario, red clump stars with high polarimetric signal-to-noise on both sides of the cloud are found and the polarization signal from the cloud can be independently measured. In the more likely scenario the polarization signal from the cloud will dominate any signals from more distant stars making analysis beyond the cloud impossible. In this case the clouds will create ``shadow zones'' where the magnetic field cannot be measured by this method.

\section{Discrete Magnetic Field Decomposition with Few Red Clump Stars}
\label{inner}

$H$-band ($1.6\mu$m) starlight polarimetry toward $(\ell , b)= (19.49\degr,+0.56\degr)$ was drawn from the first data release of the Galactic Plane Infrared Polarization Survey \citep[GPIPS;][]{GPIPS_III,GPIPS_I}. These observations were made with the Mimir instrument \citep{2007PASP..119.1385C} on the 1.8m Perkins telescope of Lowell Observatory outside Flagstaff, AZ. Each observation covered a $10'\times 10'$ field-of-view and data were reduced with custom IDL-based software \citep{GPIPS_II, GPIPS_I}. For each observation, a catalog of sky coordinates, starlight polarization properties, and measured $H$-band photometry were produced. These catalogs were matched to 2MASS and GLIMPSE \citep{2003PASP..115..953B} point source catalogs. GPIPS polarimetric observations are complete to $H\approx 12$, which provides a faintness (and ultimately distance) limit on red clump stars with measurable polarimetry. Data from the $^{13}$CO Galactic Ring Survey (Jackson et al. 2006) shows no significant emission at any velocity toward this direction indicating that it is free from cloud contamination.

\begin{figure}
	\epsscale{1.1}
	\centering
		\plotone{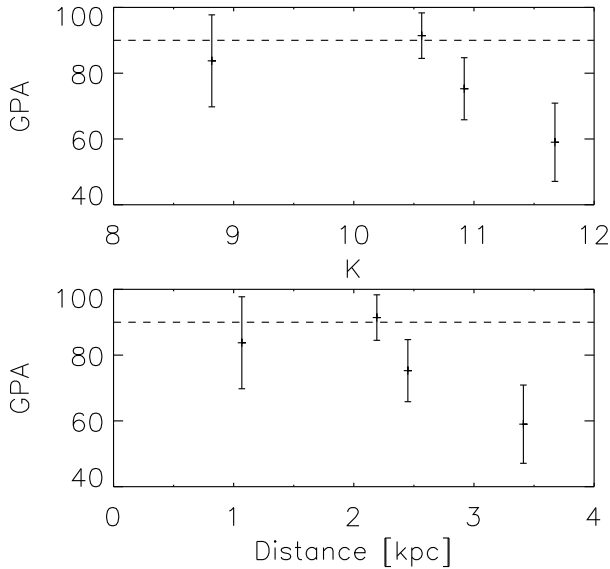}
	\caption{\label{gpa_vs_k}Top panel: near-infrared polarization orientation as Galactic position angle (GPA) as a function of 2MASS K-band apparent magnitude for red clump stars. The horizontal dashed line represents a polarization angle parallel to the Galactic disk. Bottom panel: GPA vs. estimated distances for the same stars.}
\end{figure}

Using the photometric method described in Section \ref{red_clump}, four candidate red clump stars were identified in this $10'\times 10'$ field. Using the estimated contamination fractions presented in Figure \ref{contam}, the K-band magnitude maps directly to a false-positive probability for that star. These $K$-band magnitudes are 8.82, 10.57, 10.92, and 11.67; which become false-positive probabilities of 0.141, 0.049, 0.047, and 0.045. The probability that all four stars are actually red clump stars (a reliability probability) is only 0.743 with the brightest star being the least likely. Nevertheless, the following analysis assumes that all four stars are actually red clump stars.

\begin{deluxetable*}{ccccccc}
\tablewidth{0pt}
\tablecolumns{7}
\tablecaption{Inner Galaxy $H$-band Polarimetry\label{pol_table}}
\tablehead{
	\colhead{Star} &
	\colhead{$H_{2MASS}$} &
	\colhead{Distance} &
	\colhead{P\tablenotemark{a}} &
	\colhead{GPA\tablenotemark{b}} &
	\colhead{$U$} &
	\colhead{$Q$} \\
	\colhead{} &
	\colhead{(mag)} &
	\colhead{(kpc)} &
	\colhead{(\%)} &
	\colhead{(deg)} &
	\colhead{(\%)} &
	\colhead{(\%)} \\
	\colhead{(1)} &
	\colhead{(2)} &
	\colhead{(3)} &
	\colhead{(4)} &
	\colhead{(5)} &
	\colhead{(6)} &
	\colhead{(7)} \\
	}
\startdata
1 & $9.04\pm0.02$ & $1.07\pm0.23$ & $0.31\pm0.15$ & $83.8\pm12.5$ & $0.23\pm0.15$ & $0.25\pm0.15$ \\
2 & $10.89\pm0.04$ & $2.19\pm0.61$ & $2.87\pm0.69$ & $91.4\pm  6.9$ & $2.52\pm0.68$ & $1.54\pm0.72$ \\
3 & $11.31\pm0.02$ & $2.45\pm0.56$ & $2.08\pm0.68$ & $75.3\pm  9.2$ & $0.97\pm0.71$ & $1.96\pm0.68$ \\
4 & $12.09\pm0.03$ & $3.41\pm0.87$ & $4.67\pm1.94$ & $59.0\pm11.0$ & $-0.55\pm1.94$ & $5.03\pm1.94$ \\
\enddata
\tablenotetext{a}{These P values and the derived position angle uncertainties have been corrected for positive statistical bias via the equation $P=\sqrt{U^2+Q^2-\sigma_P^2}$ \citep{1974ApJ...194..249W,2006PASP..118.1340V}.}
\tablenotetext{b}{GPA values were derived from the equatorial Stokes parameters, then rotated to the Galactic coordinate system using the equation in \citet{1968ApJ...151..907A}.}
\end{deluxetable*}

For the four red clump stars identified toward $(\ell , b)= (19.49\degr,+0.56\degr)$, the Galactic position angle (GPA) is plotted against the 2MASS $K$-band magnitude in the top panel of Figure \ref{gpa_vs_k} and all of the H-band polarization properties are listed in Table \ref{pol_table}. GPA is given in the Galactic coordinate system, where GPA=$90\degr$ is parallel to the Galactic disk, shown by the dashed line in Figure \ref{gpa_vs_k}. In the bottom panel of that Figure, the $x$-axis shows the estimated distance to each red clump star.

Qualitative estimates of the magnetic properties with distance can be deduced from Figure \ref{gpa_vs_k}. The first two stars have GPAs consistent with $90\degr$, suggesting that the magnetic field in this direction is parallel to the Galactic plane out to a distance of $\sim$2 kpc. The next two stars show a monotonic change in GPA with apparent magnitude (i.e., distance). Beyond 2 kpc, the magnetic field orientation changes. An alternative interpretation is that there is a constant change in GPA with $K$ or distance (representable by a straight line with non-zero slope) instead of a knee. A more quantitative analysis can be made using the procedure described below.

\begin{deluxetable}{ccccc}
\tablewidth{0pt}
\tablecolumns{5}
\tablecaption{Inner Galaxy Magnetic Field Properties, Binned by Distance\label{mag_table}}
\tablehead{
	\colhead{Interval\tablenotemark{a}} &
	\colhead{Distance} &
	\colhead{GPA$_i$} &
	\colhead{$\Delta U_i$} &
	\colhead{$\Delta Q_i$} \\
	\colhead{} &
	\colhead{Range} &
	\colhead{} &
	\colhead{} &
	\colhead{} \\
	\colhead{} &
	\colhead{[kpc]} &
	\colhead{[deg]} &
	\colhead{[\%]} &
	\colhead{[\%]} \\
	\colhead{(1)} &
	\colhead{(2)} &
	\colhead{(3)} &
	\colhead{(4)} &
	\colhead{(5)} \\
	}
\startdata
0$\rightarrow$1 &  0$\rightarrow$1.07      & $83.8\pm12.5$ & $0.23\pm0.15$ & $0.25\pm0.15$  \\
1$\rightarrow$2 &  1.07$\rightarrow$2.19 & $92.4\pm7.9$   & $2.29\pm0.70$ & $1.29\pm0.74$  \\
2$\rightarrow$3 &  2.19$\rightarrow$2.45 & $24.7\pm17.6$ & $-1.55\pm0.98$ & $0.42\pm0.99$  \\
3$\rightarrow$4 &  2.45$\rightarrow$3.41 & $49.0\pm17.3$ & $-1.52\pm2.07$ & $3.07\pm2.06$  \\
\enddata
\tablenotetext{a}{These numbers refer to the stars listed in Table \ref{pol_table}, zero being the observer.}
\end{deluxetable}

The properties of the polarizing layers toward this direction were derived using Stokes subtraction. Table \ref{mag_table} lists the derived properties for the polarizing intervals between each adjacent pairs of stars. Each interval is designated by the star identifiers in Table \ref{pol_table}, zero referring to an observer on Earth. The distance span of each interval is bounded by the estimated distances of the stars, zero for an observer. The differences in Stokes parameters for the stars bounding each interval, $\Delta U_i$ and $\Delta Q_i$, and the corresponding polarization P.A. (rotated from equatorial to the Galactic coordinate system) are also listed. The corresponding degree of polarization is not listed because this quantity correlates more with the intervening dust column than with the magnetic field properties \citep{2011ApJ...740...21P}.

\begin{figure}
	\epsscale{1.1}
	\centering
		\plotone{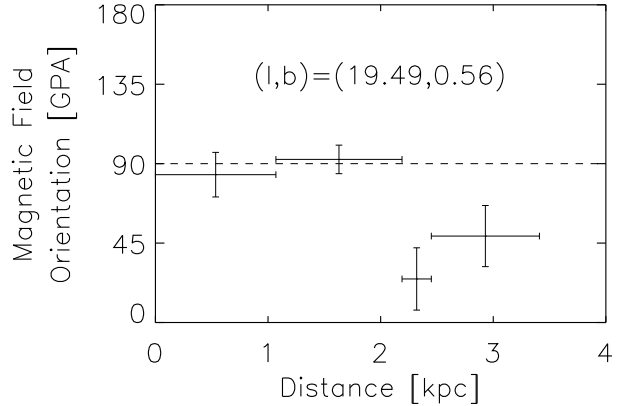}
	\caption{\label{mag_vs_dist}Sky-projected magnetic field orientation as a function of distance toward $(\ell , \it{b})$ = (19.49\degr,+0.56\degr). The horizontal error bar represents the range over which this estimate spans, and the vertical error bar is the uncertainty in the sky-projected magnetic field orientation for that distance span. The horizontal dashed line shows GPA=$90\degr$, which represents alignment with the Galactic plane.}
\end{figure}

The derived dependence of the magnetic field orientation with distance along this direction is shown in Figure \ref{mag_vs_dist}. There, the horizontal error bars represent the distance intervals listed in Table \ref{mag_table} and the vertical error bars are the uncertainty in the average, dust-weighted magnetic field orientation in that interval. Out to a distance of 2 kpc, the magnetic field orientation is consistent with GPA$=90\degr$. Beyond 2 kpc, the magnetic field orientation is significantly different, being tilted approximately $45\degr$ to the direction of the Galactic plane.

From the $^{13}$CO data, it is unlikely that there is a cloud along this particular line of sight that could create a false signal. To further test this possibility the extinction per unit length (A$_v$/L) for the four stars toward this direction was calculated. If there was a dense cloud near 2 kpc there should be a large jump in A$_v$/L beyond that distance. A$_v$/L for the four stars are (in order of increasing distance): 1.88, 1.90, 1.67, and 1.46. The lack of a jump in A$_v$/L is further evidence against the presence of an intervening cloud. In fact, this result suggests that there is less extinction at further distances (probably because of increasing Galactic height).

This result is somewhat robust against the individual false-positive probabilities calculated above. The removal of any one star may increase the distance uncertainty of the change in magnetic orientation, but it would still remain detected. For example, the brightest star in Figure \ref{mag_vs_dist} has the lowest probability of being a red clump star. Excluding this star from the previous analysis does not affect the presence of the change in magnetic field orientation beyond 2 kpc, but actually increases the red clump reliability from 0.743 to 0.865.

\section{Continuous Magnetic Field Decomposition with Many Red Clump Stars}
\label{outer}

An inherent drawback of this method, well demonstrated toward the inner Galaxy line of sight in the previous section, is the small number of identifiable red clump stars with measured NIR polarization in each $10'\times 10'$ GPIPS field. With so few stars, measurement errors could easily mimic a false signal and impede interpretation of the results, any intrinsic stellar polarization could also mask the signature of the Galactic magnetic field, and false detections would introduce noise into the results. Furthermore, unidentified dense regions along the line of sight can significantly affect interpretation of the measured polarizations. Larger numbers of red clump stars were available in a sample of starlight polarimetry drawn from \citet{2012ApJ...749...71P} and is here used to illustrate this method toward an example direction in the outer Galaxy, $(\ell , \it{b})$ = (150\degr, -2.5\degr). 

\begin{figure}
	\epsscale{1.1}
	\centering
		\plotone{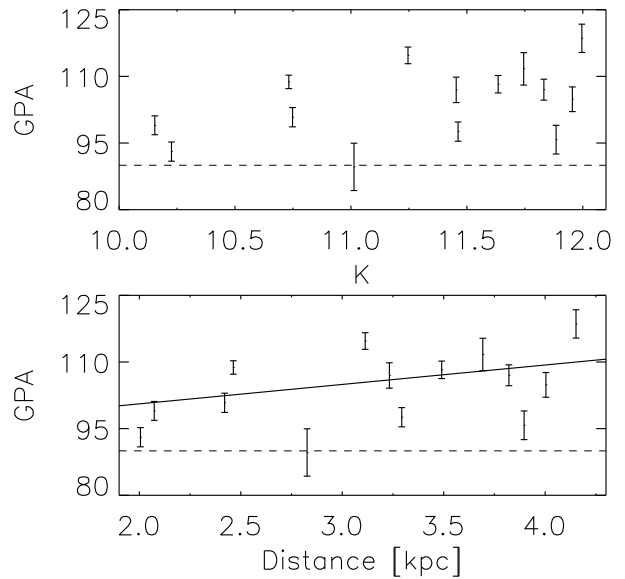}
	\caption{\label{outer_gal}Same as Figure \ref{gpa_vs_k}, but toward the $10' \times 10'$ field centered on $(\ell , \it{b})$ = (150\degr, -2.5\degr).}
\end{figure}

Red clump stars were identified in this direction using the same procedure described in Section \ref{red_clump}, and many more stars were found than in the previous inner Galaxy example. The larger number of red clump stars found in this outer Galaxy example is likely caused by lower extinction allowing larger distances to be probed, which can be seen in the bottom panels of Figures \ref{gpa_vs_k} and \ref{outer_gal}. Plots of the measured polarization GPA with apparent magnitude and estimated distance are shown in Figure \ref{outer_gal}. 

\begin{figure}
	\epsscale{1.1}
	\centering
		\plotone{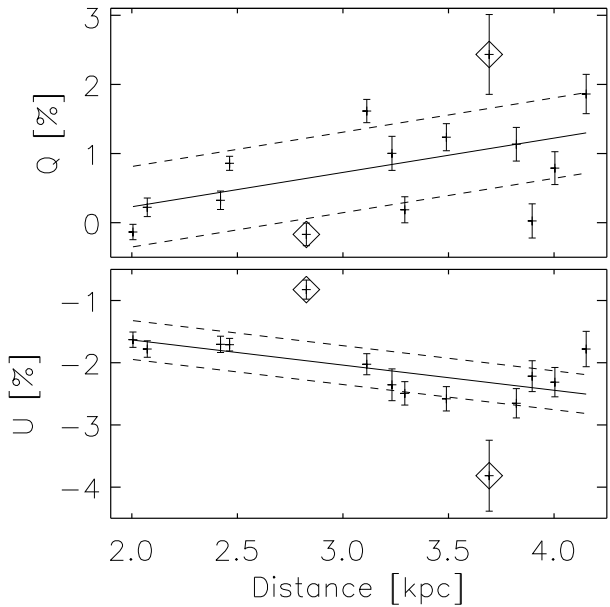}
	\caption{\label{outer_gal_uq}Stokes $Q$ (top panel) and $U$ (bottom panel) as a function of distance for red clump stars toward the $(\ell , \it{b})$ = (150\degr, -2.5\degr) field. The solid lines are linear fits to the data, the dashed lines show the standard error of that fit. Stars with possible intrinsic polarization (marked with diamonds) were excluded from the fits.}
\end{figure}

The larger number of stars available toward this outer Galaxy line of sight allows a more sophisticated analysis of changes in magnetic field orientation with distance. By fitting the change in Stokes parameters with estimated distance, the geometry of the sky-projected magnetic field can be measured without needing to invoke a specific model. Figure \ref{outer_gal_uq} shows the variation of Stokes $U$ and $Q$ parameters with estimated distance. Excluding the two stars showing evidence for intrinsic polarization (discussed below), a least-squares fit to the data was calculated and is also plotted as solid lines. A statistical F-test showed that a linear fit was appropriate and that higher order fitting terms were insignificant in both cases. In Section \ref{inner}, the entire line of sight was divided into four regions, here the sampled line of sight is divided into infinitesimally small regions and the instantaneous change in Stokes parameters is calculated from the fits in Figure \ref{outer_gal_uq}. The change in Stokes parameters with distance maps directly to the orientation of the magnetic field via the equation:
\begin{equation}
{GPA_i}=\frac{1}{2} \, atan \left[ \frac{dU_i/ds}{dQ_i/ds} \right]
\end{equation}
where $i$ refers to the $i$th polarizing region along the line of sight and $ds$ is an infinitesimal change in distance along this direction. For the fits in Figure \ref{outer_gal_uq}, $dU/ds=-0.404 \pm 0.312$ kpc$^{-1}$ and $dQ/ds=0.497 \pm 0.582$ kpc$^{-1}$. Since these values are constants, this is interpreted as a constant magnetic field orientation in this direction between 2 and 5 kpc with equatorial PA$=160.5\degr \pm 19.6\degr$ and GPA$=120.9\degr \pm 19.6\degr$.

This result can be seen in Figure \ref{outer_gal}. The line of sight between the observer and the first star at 2 kpc imposes a net polarization with GPA$=99.0\degr \pm 2.1\degr$. Stars at larger distances show a weak monotonic increase in P.A. toward GPA$=118.6 \pm 3.2\degr$. The polarization of a star at a distance of 5 kpc is dominated by the polarizing layer between 2 and 5 kpc with a constant mean magnetic field orientation measured by the slopes of $dU/ds$ and $dQ/ds$. 

\section{Identification of Intrinsically Polarized Stars}

The application of this method assumes that the stars are intrinsically unpolarized. Any stars violating this assumption could be interpreted as changes in the magnetic field geometry in the polarizing medium between the stars. This could be particularly troublesome for the `sparse' example in Section \ref{inner} where the polarization of each star was used to measure the magnetic field properties between adjacent stars. The `rich' example in Section \ref{outer} uses a statistically more robust sample to fit the magnetic field properties with distance. The polarization properties of this more robust sample can be used to tentatively identify intrinsically polarized stars by identifying significant polarization deviants toward a given direction.

\begin{figure}
	\epsscale{1.1}
	\centering
		\plotone{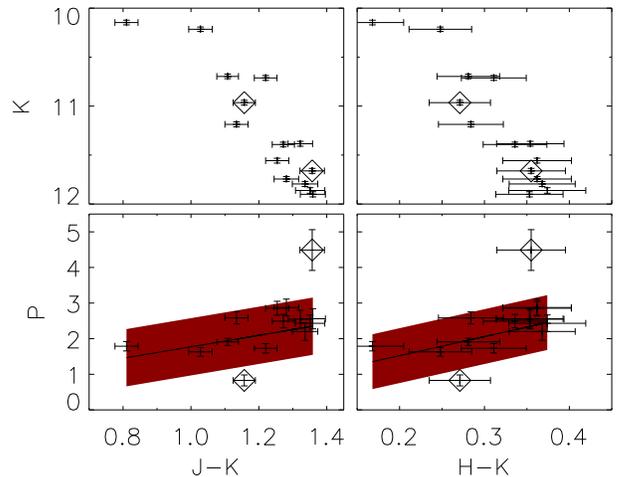}
	\caption{\label{intrinsic}Photometric method for identifying candidate red clump stars with intrinsic polarization. Lower panels show the degree of H-band polarization against $(J-K)$ (lower left panel) and $(H-K)$ (lower right panel) colors for outer Galaxy red clump stars. The solid lines are linear fits to each data set and the grey bands represents the standard errors of the fits. Stars outside this band tentatively show anomalous polarizations. Upper panels show near-infrared color-magnitude diagrams for these red clump stars, the intrinsically polarized stars are marked with diamonds. (A color version of this figure is available in the online journal.)}
\end{figure}

The 2MASS photometry and $H$-band degree of polarization of these Outer Galaxy red clump stars are shown in Figure \ref{intrinsic}. In the figure, the top panels show two CMDs: $K$ versus $J-K$ (upper left panel) and $K$ versus $H-K$ (upper right panel). If they are truly red clump stars then all should have the same absolute $K$-band magnitude and intrinsic colors. The overall trend in each CMD illustrates the red clump selection method in Section \ref{red_clump}: more distant standard candles are fainter and redder. The lower panels plot the degree of $H$-band polarization against these same colors. In these lower panels, a linear fit to the data is shown by the solid line, which recovers the expected polarization properties for the magnetic field derived in Section \ref{outer}, and the shaded regions represents the area enclosed by the standard error of the fit. In both bottom panels, the same two outliers are seen outside of the shaded region.

Both polarization and extinction (traced by NIR color) are integral quantities along the line of sight, meaning that the observed properties depend on the conditions between the object and the observer. While extinction monotonically increases with distance, the Stokes $U$ and $Q$ parameters are signed quantities, meaning that the degree of polarization is not necessarily a monotonic function of distance. In Figure \ref{intrinsic}, the photometric properties of the stars are `well-behaved' in that there is a correlation between apparent $K$ magnitude and color, because both fundamentally depend on distance. For a line of sight with a constant magnetic field orientation, the degree of starlight polarization would also depend monotonically on distance, and therefore would also correlate with apparent brightness and reddening.

The degree of polarization has a more complicated dependence on distance if the line of sight suffers depolarization \citep{1989ApJ...346..728J, 1992ApJ...389..602J}. Strong depolarization, for example occurring as a result of a change in the magnetic field direction at some distance, would cause the polarization trend to change and all stars beyond that distance would be affected. Similarly, the presence of a polarizing cloud along the line of sight would also affect all stars beyond its distance. This would not give rise to individual outliers, but to a discreet offset in polarization versus distance at that distance. A more likely explanation for individual outliers is that some intrinsic property of the star is contributing to its measured polarization.

The variation in Stokes $U$ and $Q$ parameters with $K$ are shown in Figure \ref{outer_gal_uq} and can distinguish between intrinsically polarized stars and the presence of a change in field orientation or a polarizing cloud. In that figure the candidate intrinsically polarized red clump stars are marked with diamonds. A linear fit to the data, excluding the intrinsically polarized candidates, is shown by the solid line. The dashed lines represent the standard error of that fit. If there were a systematic change in the magnetic field orientation or a polarizing cloud along the line of sight there would be bend or offset in the trend of $U$ and $Q$ with distance.

The candidate stars appear as outliers in that figure as well, particularly in Stokes $U$ (lower panel). The intrinsic polarization of each star (assuming they are not interlopers) was estimated by the difference between the fit and the observed Stokes parameters. For the brighter candidate star at $K=11.01$ mag (2MASS J03534501+5026427), the intrinsic degree of polarization is $1.58\%\pm0.45\%$ with GPA$=114.0\pm14.9\degr$. The fainter candidate star (2MASS J03543283+5029199) at $K=11.75$ mag has an intrinsic polarization of $1.94\%\pm0.73\%$  with GPA$=116.3\pm21.4\degr$.

These stars were assumed to be red clump stars. However, they could be true interlopers in which case their distance estimates would be incorrect. If they were distant reddened supergiants or dwarfs then the measured polarizations could be consistent with the rest of that line of sight. No additional information is currently available in the literature for these two objects, so their true identity or the cause of this anomalous polarization remains unknown. Spectroscopic follow-up could easily determine whether these stars belong to the red clump or not. Some possible explanations for anomalous polarization of red clump stars are presented below.

\section{Discussion}

Red clump stars were used as standard candles to probe the structure of the sky-projected magnetic field along two Galactic lines-of-sight. toward one field in the inner Galaxy, the magnetic field showed evidence for a large change in magnetic field orientation at a specific distance. toward the outer Galaxy, the magnetic field orientation was uniform in the region probed, however not parallel to the Galactic disk.

\subsection{Evidence for a Magnetic Reversal?}

In Figure \ref{mag_vs_dist}, the H-band polarimetry toward the inner Galaxy showed a change in the magnetic field orientation at a distance of $\sim$2 kpc. Apparently, the sky-projected magnetic field orientation has changed, but this could also be evidence for the presence of magnetic reversals in the Galactic magnetic field.

Considering the Galactic center located at the origin of a cylindrical coordinate system, the magnetic field at any point can be defined by the vector sum of radial, vertical, and azimuthal components. In the disk, theoretically \citep[e.g., ][and references therein]{1978mfge.book.....M, 2000AA...358..125F, 2008RPPh...71d6901K} and observationally \citep[e.g., ][and references therein]{1976ARAA..14....1H, 1996ARAA..34..155B}, the azimuthal component dominates the total field strength in the disk of spiral galaxies, including the Milky Way.

In the Milky Way, the dominant azimuthal component of the magnetic field does not necessarily have the same sign everywhere. Evidence for discreet azimuthal magnetic field reversals have been seen in the Galactic disk in Faraday rotation measures \citep[RMs;][]{2008AA...477..573S,2011ApJ...728...97V}. As described in \citet{2011ApJ...740...21P}, starlight polarimetry is insensitive to magnetic field direction, only its orientation, leading to a $180\degr$ ambiguity in the direction of the magnetic field aligning the dust. Formally, starlight polarimetry should be insensitive to full $180\degr$ magnetic reversals. However, if instead of simply reversing the azimuthal magnetic field direction, the azimuthal reversal is associated with a large-scale perturbation in the vertical and/or radial magnetic field components, then the projection of that field onto the sky would also change and the effects of magnetic reversals would be expected to be seen with starlight polarimetry.

Magnetic reversals are expected in two main situations: (1) there is a field discontinuity, separated by a current sheet \citep{1954ApJ...119....1F}, or (2) if the magnetic field is continuous, then the amplitude of the azimuthal magnetic field must decrease to zero before increasing in the opposite direction, though no limits are placed on the vertical and radial field components that may also change, possibly in such a way that the total field strength is constant. In the first situation the magnetic field direction is reversed on opposite sides of a current sheet. Magnetic field rotations of $180\degr$ have been observed by in situ satellites passing through the heliospheric current sheet in the solar system \citep{2004JGRA..109.3107C}. The presence of a current sheet would have little effect on the observed polarization orientation since the magnetic fields on either side of an interstellar current sheet would be indistinguishable with this  NIR background starlight observational tool, illustrated in Figure 1 of \citet{2011ApJ...740...21P}. In the second situation, as the dominant azimuthal magnetic field vanishes the other field components (radial and vertical) will be expressed. This would manifest as a change in the linear polarizing properties of the medium hosting this magnetic field.

A change in the magnetic field geometry toward the inner Galaxy has been modeled from observations of RMs of Galactic pulsars and polarized extragalactic objects. toward the direction shown in Figure \ref{mag_vs_dist} ($\ell=20\degr$, $b=0\degr$), \citet{2008AA...477..573S} saw evidence for magnetic reversals at Galactocentric radii (R) of 7.5, 6, and 5 kpc (corresponding to line of sight distances of 1.1, 2.7, and 3.9 kpc, respectively) in their ASS+RING Galactic magnetic field model. In this same direction \citet{2011ApJ...728...97V} concluded that a magnetic reversal occurs at R=5.8 kpc (a line of sight distance of 3.0 kpc). It seems that the same magnetic feature seen in the RMs is being observed by NIR starlight polarimetry. If so, this would corroborate the presence of a large-scale magnetic field change toward that direction.

For now, the NIR starlight polarimetry results, while intriguing, are tentative. Spectroscopic follow-up of the stars along this line of sight is required to derive reliable spectrophotometric distances, but will require a substantial observational program to test the photometric results presented here.

\subsection{Magnetic Fields Following the Galactic Warp}

Figure \ref{outer_gal} shows a line of sight toward ($\ell$, $b$) = ($150\degr$, $-2.5\degr$). Toward this direction in the outer Galaxy, the magnetic field orientation was shown to be uniform with distance, which agrees with the observationally-driven models of \citet{2008AA...477..573S} and \citet{2011ApJ...728...97V}. Observations and simulations conclude that in this direction the disk magnetic field should be roughly parallel the Galactic plane \citep{1996ARAA..34..155B,2009ASTRA...5...43B}.

However, the deviation of the polarization P.A., as described in Section \ref{outer} (GPA$=121\degr$), is inconsistent with a plane-parallel magnetic field in the Galactic disk (GPA$=90\degr$). Geometric effects cannot account for the discrepancy. This particular outer Galaxy line of sight samples the Galactic magnetic field below the disk ($b=-2.5$), and the combination of inclined viewing angle \citep{2011ApJ...740...21P} and the pitch angle of the Galactic disk magnetic field \citep{1988AJ.....95..750V, 1994AA...288..759H, 1996ApJ...462..316H, 2007EAS....23...19B, 2011ApJ...738..192P, 2012ApJ...749...71P} only contribute a 1.1$\degr$ twist to the projected magnetic field orientation for stars within $\sim 7$ kpc.

A possible explanation for the observed magnetic field orientation in the Outer Galaxy is the presence of the Galactic warp . Using the model of \citet{2002AA...394..883L}, the Galactic warp can be described by $z(\phi)\sim sin(\phi+5\degr)$ where $\phi = 0\degr$ is toward the Sun and $\ell=150\degr$ spans $\phi=0-12.3\degr$ from the Sun to a distance of 4 kpc. For these values of $\phi$, the Galactic plane is tilted upward $44\degr$ toward Galactic North. This leads to a predicted magnetic field orientation of GPA=$134\degr$, consistent with the measured value (GPA$=121\pm20\degr$). A similar result was seen in M33 where vertical magnetic fields were suggested to actually be disk fields perturbed by a Galactic warp \citep{2008A&A...490.1005T}.

\subsection{Intrinsically Polarized Red Clump Stars}

One unexpected result of this study was the possible identification of intrinsically polarized red clump star candidates. These stars were photometrically identified as red clump stars, but their polarization properties (shown in Figures \ref{intrinsic} and \ref{outer_gal_uq}) were inconsistent with the trend of Stokes $U$ and $Q$ versus brightness (or distance) seen in the other red clump stars. The reason for intrinsic polarization is unclear from the data available, though there are several possibilities. Rapid rotation in old red clump stars is unlikely, but its presence could distort the star from spherical symmetry leading to a net polarization signature \citep[up to 1.7\%;][]{1968ApJ...151.1051H}. This rotation would be easily seen in moderate resolution spectroscopy and also allow identification of the star's rotation axis. The presence of a disk or envelope could also give rise to a net polarization \citep[up to 4\%;][]{1996ApJ...461..828W}. If true, the observed NIR colors would suggest that this is not a red clump star but possibly a reddened main sequence star that is observed in the same color-magnitude space as true red clump stars. Future variability studies or spectroscopy could shed light on the physical mechanism creating these apparently anomalous polarizations.

The method described above can be used to measure the mean interstellar polarization signal with distance for some direction, then easily distinguish polarimetric deviants from stars following the nominal polarization trend. This allows for the possibility of surveying for intrinsically polarized stars over large areas of the sky. Intrinsic polarization is associated with a range of physical phenomena that may not otherwise be seen in large-scale photometric surveys. By looking for polarization deviants, a catalog of interesting targets can be developed for follow-up investigation.

\section{Summary and Conclusions}

This work presents a new method for measuring the structure of the large-scale Galactic magnetic field. Photometrically identified red clump stars were used as standard candles along two example directions in the sky, one with few red clump stars and one with many red clump stars. The false-detection probability as a function of $K$-band brightness when using the method of \citet{2002AA...394..883L} was calculated and used to inform the reliability of the results. H-band starlight polarimetry of these stars from GPIPS provides information about the integrated magnetic field properties toward each star. By decomposing the Stokes $U$ and $Q$ parameters with distance, changes in the magnetic field orientation in the plane of the sky were measured with distance.

One line of sight, toward $(\ell , b)= (19.49\degr,+0.56\degr)$ in the inner Galaxy, shows evidence for a strong magnetic field perturbation at a distance of 2 kpc that cannot be attributed to dusty clouds along the line of sight. This is approximately the same distance as a magnetic field reversal seen in Faraday rotation surveys \citep{2008AA...477..573S, 2011ApJ...728...97V}, which are sensitive to the electron-weighted magnetic field. Formally, starlight polarimetry should not be sensitive to magnetic field reversals \citep{2011ApJ...740...21P}, but changes in large-scale magnetic field geometry could manifest as both NIR polarization P.A. changes and as magnetic field reversals in Faraday rotation studies. While this remains circumstantial evidence for the presence of a magnetic field reversal toward this direction, the overall agreement between Faraday rotation studies of the Galactic magnetic field and this method is intriguing.

A different line of sight, toward $(\ell , b)= (150\degr,-2.5\degr)$ in the outer Galaxy, shows a constant magnetic field orientation between distances of 2-5 kpc. The larger number of identified red clump stars allowed a more detailed analysis of the relatively simple magnetic field in this direction. The measured magnetic field orientation (GPA=$121\degr$) does not fit the plane-parallel model typically assumed for the Galactic plane. However, this value can be explained by the Galactic warp toward this direction in the Galaxy. From this analysis, it appears that the large-scale Galactic magnetic field toward this direction follows the Galactic warp.

An unexpected outcome of the analysis toward the outer Galaxy was the possible detection of two intrinsically polarized red clump stars. Photometrically, these stars fall along the red clump sequence, but their polarization properties do not match the growth of percentage polarization with distance exhibited by the remaining stars. The deviations of the NIR polarization observations of these stars to the fits of the other stars' Stokes parameters allowed determination of these intrinsic polarizations. Moderate resolution spectra (to test whether these are red clump stars and to measure $v$ sin $i$) would provide additional insight into the physical natures of these objects.

This work represents a first step in decomposing with distance the magnetic fields probed by background starlight polarimetry. Additional work is required to fully develop this method. Photometric methods can statistically identify red clump stars; however, spectroscopic follow-up is needed. Also, while the relative distances, or at least distance order, of stars derived from photometry are reliable, physical distances require spectrophotometric follow-up. The future application of photometrically identified red clump stars to large, uniformly sampled background starlight polarimetry will allow the large-scale Galactic magnetic field to be unambiguously disentangled with distance for the first time.

\acknowledgements
The author would like to thank Dan P. Clemens, Brian W. Walsh, Kamen Kozarev, Daniel Jaffe, and Neal Evans for useful discussions. The author would also like to thank the anonymous referee for comments that greatly improved the manuscript. The author gratefully acknowledges financial support from McDonald Observatory. This research used data from the Boston University (BU) Galactic Plane Infrared Polarization Survey (GPIPS), funded in part by NSF grants AST 06-07500 and 09-07790. GPIPS used the Mimir instrument, jointly developed at BU and Lowell Observatory and supported by NASA, NSF, and the W.M. Keck Foundation.

\end{document}